\documentclass[aps,pra,twocolumn]{revtex4}
\usepackage{graphicx}
\usepackage{dcolumn}
\usepackage{amsmath}
\usepackage{amssymb}
\usepackage{float}
\usepackage{epstopdf}

\def\itGa{{\mathit{\Gamma}}}

\def\itOm{{\mathit{\Omega}}}

\def\rel{_{\rm rel}}
\def\cm{_{\rm cm}}

\def\ext{_{\rm ext}}

\def\xw{w}

\def\har{_{\rm H}}
\def\ee{_{\rm ee}}

\def\mdl{^{\rm AR}}

\def\cor{_{\rm c}}
\def\glt{^{\rm GL2}}

\def\xc{_{\rm xc}}
\def\x{_{\rm x}}
\def\c{_{\rm c}}
\def\e{_{\rm e}}

\def\rcm{{\bf R}}
\def\rv{{\bf r}}

\def\sv{{\bf s}}

\def\beq{\begin{equation}}
\def\eeq{\end{equation}}

\begin{document}
\title{Adiabatic connection at negative coupling strengths}
\author{Michael Seidl$^1$ and Paola Gori-Giorgi$^{2,3}$}
\affiliation{$^1$Institute of Theoretical Physics,
University of Regensburg, D-93040 Regensburg, Germany \\
$^2$Laboratoire de Chimie Th\'eorique, CNRS,
Universit\'e Pierre et Marie Curie, 4 Place Jussieu, F-75252 Paris, France \\
$^3$Theoretische Chemie, Vrije Universiteit Amsterdam, De Boelelaan 1083, 1081HV Amsterdam, The Netherlands}

\date{\today}

\begin{abstract}

The adiabatic connection of density functional theory (DFT) for electronic systems is generalized here to negative values of the coupling strength $\alpha$ (with {\em attractive} electrons). In the extreme limit $\alpha\to-\infty$ a simple physical solution is presented and its implications for DFT (as well as its limitations) are discussed. For two-electron systems (a case in which the present solution can be calculated exactly), we find that an interpolation between the limit $\alpha\to-\infty$ and the opposite limit of infinitely strong repulsion ($\alpha\to+\infty$) yields a rather accurate estimate of the second-order correlation energy $E\cor\glt[\rho]$ for several different densities $\rho$, without using virtual orbitals. The same procedure is also applied to the Be isoelectronic series, analyzing the effects of near-degeneracy. 
\end{abstract}

\maketitle
\section{Introduction, definitions and outline}

Combining low computational cost with reasonable accuracy for many molecules and solids, density functional theory (DFT) \cite{ParYan-BOOK-89,Koh-RMP-99} has become a particularly successful approach for electronic-structure calculations, both in chemistry and in physics.
In DFT, the exact electron density $\rho=\rho(\rv)$ and ground-state energy $E_N[v]$ of $N$ interacting electrons in a given external potential $v=v(\rv)$ could be in principle obtained by
solving single-particle (``Kohn-Sham'') equations for non-interacting electrons. In the practical implementation of Kohn-Sham (KS) DFT, however, we have to rely on approximations for
the density functional $E\xc[\rho]$ of the exchange-correlation (xc) energy.
Despite the large number of available approximations for this functional and of their successful applications, there are still important cases in which KS DFT can fail, which is why the quest for better xc functionals continues to be a very active research field (for recent reviews see, e.g., Refs.~\onlinecite{Mat-SCI-02,PerRuzTaoStaScuCso-JCP-05,CohMorYan-SCI-08}). For example, present-days KS DFT encounters problems in the treatment of near-degeneracy effects (rearrangement of electrons within partially filled levels, important for describing bond dissociation but also equilibrium geometries, particularly for systems with $d$ and $f$ unsaturated shells), in the description of van der Waals long-range interactions (relevant, for example, for biomolecules and layered materials),  and of localization effects due to strong electronic correlations, as those occurring in Mott insulators or in low-density nanodevices. 

An exact expression for the xc functional is the coupling-constant integral 
\cite{HarJon-JPF-74,LanPer-SSC-75,GunLun-PRB-76},
\beq
E\xc[\rho]=\int_0^1d\alpha W_\alpha[\rho].
\label{CCI}
\eeq
The integrand is the $\alpha$-dependent density functional
\beq
W_\alpha[\rho]=\langle\Psi_\alpha[\rho]|\hat{V}\ee|\Psi_\alpha[\rho]\rangle
                                     -U[\rho],
\label{Walpha}
\eeq
with the operator of the electronic Coulomb repulsion,
\beq
\hat{V}\ee=\frac{e^2}2\sum_{i=1}^N\sum_{j=1}^N\frac{1-\delta_{ij}}{|\rv_i-\rv_j|},
\label{Vee}
\eeq
and the continuum functional of the Hartree energy,
\beq
U[\rho]=\frac{e^2}2\int d^3r\int d^3r'\frac{\rho(\rv)\rho(\rv')}{|\rv-\rv'|}.
\eeq
The crucial quantity in Eq.~\eqref{Walpha} is the $\alpha$-dependent wave
function
\beq
\Psi_\alpha[\rho]=\Psi_\alpha([\rho];\rv_1,...,\rv_N;\sigma_1,...,\sigma_N),
\eeq
where the $\rv_n$ and the $\sigma_n$, respectively, are spatial and spin coordinates
of the electrons. Out of all antisymmetric
$N$-electron wave functions $\Psi$ that are associated with the same given
electron density $\rho$, $\Psi_{\alpha}[\rho]$ denotes the one that
yields the minimum expectation of $\hat{T}+\alpha\hat{V}\ee$ \cite{Lev-PNAS-79}, with the
kinetic-energy operator $\hat{T}=-\frac{\hbar^2}{2m\e}\sum_{i=1}^{N}\nabla_i^2$,
\beq
\langle\Psi_{\alpha}[\rho]|\hat{T}+\alpha\hat{V}\ee|\Psi_{\alpha}[\rho]\rangle
=\min_{\Psi\to\rho}\langle\Psi|\hat{T}+\alpha\hat{V}\ee|\Psi\rangle.
\eeq
If the density $\rho$ is $v$-representable for all $\alpha\ge 0$, there
exists an $\alpha$-dependent external potential $v\ext^\alpha([\rho],\rv)$
such that $\Psi_{\alpha}[\rho]$ is the ground state of the Hamiltonian
\beq
\hat{H}_\alpha[\rho]=\hat{T}+\alpha\hat{V}\ee
                +\sum_{i=1}^Nv\ext^\alpha([\rho],\rv_i).
\label{Hv}
\eeq
By construction, this Hamiltonian has for all $\alpha\geq0$ the same
ground-state density $\rho=\rho(\rv)$ as the real system with $\alpha=1$.

At the non-interacting limit $\alpha=0$, the ground state of
$\hat{H}_\alpha[\rho]$ is, in most cases, the single Slater determinant $\Psi_0[\rho]$ with the $N$
occupied Kohn-Sham orbitals. Consequently,
\beq
E\x[\rho]\equiv W_0[\rho]=\langle\Psi_0[\rho]|\hat{V}\ee|\Psi_0[\rho]\rangle
                         -U[\rho]
\label{Ex}
\eeq
is the functional of the DFT exchange energy. 

For $\alpha>0$,
$\Psi_{\alpha}[\rho]$ is no longer a Slater determinant of single-particle
orbitals, but a correlated $N$-electron wave function.
As $\alpha>0$ grows, the electron-electron repulsion in the state
$\Psi_{\alpha}[\rho]$ increases. So does the average distance 
$\langle|\rv_i-\rv_j|\rangle$ between two electrons. Consequently, for
$N>1$,  the expectation of $\hat{V}\ee$ in Eq.~\eqref{Walpha} must be a monotonically
decreasing function of $\alpha$, $\frac{d}{d\alpha}W_\alpha[\rho]<0$.

The quantity
\beq
E\c\glt[\rho]\equiv\frac12\frac{d}{d\alpha}W_\alpha[\rho]\Big|_{\alpha=0}
\eeq
is the 2nd-order correlation energy in the G\"orling-Levy perturbation expansion
\cite{GorLev-PRB-93,GorLev-PRA-94}. It can be expressed in terms of the KS single-particle orbitals,
but, in contrast to $E\x[\rho]$, it requires also all the unoccupied orbitals,
\beq
E\c\glt[\rho]=-\sum\limits _{k=1}^{\infty}
\frac{|\langle\Psi_0^k[\rho]|\;\hat{V}\ee\!-\!
\hat{V}\har[\rho]\!-\!\hat{V}\x[\rho]\;|\Psi _0[\rho]\rangle |^2}{E_0^k -E_0}.
\label{EcGL2}
\eeq
Here, $\Psi_0^k[\rho]$ is the $k$-th excited state and $E_0^k$ the
corresponding eigenvalue (while $E_0$ is the ground-state energy) of the
non-interacting Hamiltonian $\hat{H}_0[\rho]$.
The operators $\hat{V}\har[\rho]=\sum_{i=1}^Nv\har([\rho];\rv_i)$ and
$\hat{V}\x[\rho]=\sum_{i=1}^Nv\x([\rho];\rv_i)$, respectively, represent
the Hartree potential,
\beq
v\har([\rho];\rv)=\frac{\delta U[\rho]}{\delta\rho(\rv)}\equiv
e^2\int d^3r'\frac{\rho(\rv')}{|\rv-\rv'|},
\eeq
and the exchange potential,
\beq
v\x([\rho];\rv)=\frac{\delta E\x[\rho]}{\delta\rho(\rv)}.
\eeq
Since the exchange functional $E\x[\rho]$ is not known explicitly in terms of
the density $\rho$, but only implicitly via the KS orbitals [Eq.~\eqref{Ex}], the
evaluation of the function $v\x([\rho];\rv)$ for a given density
$\rho$ is a non-trivial problem (see, e.g., \cite{StaScuDav-JCP-06,RohGriBae-CPL-06,HesGotDelGor-JCP-07,IzmStaScuDavStoCan-JCP-07,GorHesJonLev-JCP-08,HeaYan-JCP-08}). The resulting weak-interaction expansion of $W_\alpha[\rho]$ is then
\beq
W_\alpha[\rho]=E\x[\rho]+2E\c\glt[\rho]\alpha+...\quad(\alpha\to0).
\label{wil}
\eeq

The functional $W_\alpha[\rho]$ and its exact properties have always played a central role for the construction of approximate $E\xc[\rho]$ (see, e.g., \cite{Bec-JCP-93a,Bec-JCP-93,Ern-CPL-96,BurErnPer-CPL-97,MorCohYan-JCPa-06}).
Although the integration over $\alpha$ in Eq.~(\ref{CCI}) runs between 0 and 1, we can consider values  of $\alpha$ larger than the physical interaction strength, $\alpha>1$. In particular, the strong-interaction (or low-density) limit of DFT is defined as the $\alpha\to\infty$ limit of $W_\alpha[\rho]$. It has been shown that in this limit the leading terms in $W_\alpha[\rho]$ are \cite{SeiPerLev-PRA-99,Sei-PRA-99,SeiPerKur-PRA-00,SeiGorSav-PRA-07,GorVigSei-JCTC-09},
\beq
W_\alpha[\rho]=W_\infty[\rho]+\frac{W'_\infty[\rho]}{\sqrt{\alpha}}+
\frac{W''_\infty[\rho]}{\alpha}+...\quad
(\alpha\to\infty).
\label{sil}
\eeq
While generally $W''_\infty[\rho]=0$ \cite{SeiPerKur-PRA-00,GorVigSei-JCTC-09,LiuBur-PRA-09},
the coefficients $W_\infty[\rho]$ and $W'_\infty[\rho]$ have been evaluated systematically
for spherically symmetric $N$-electron densities \cite{SeiGorSav-PRA-07,GorVigSei-JCTC-09}.
This $\alpha\to\infty$ expansion of $W_\alpha[\rho]$ is useful in several ways. For example (even if treated in an approximate way \cite{SeiPerKur-PRA-00}), it has been used to build an exchange correlation functional (interaction-strength interpolation, or ISI) by interpolating $W_\alpha[\rho]$  between $\alpha\to 0$ [Eq.~(\ref{wil})] and $\alpha\to\infty$, yielding atomization energies of simple molecules with errors within $\sim 3.4\;{\rm kcal/mol}$ \cite{SeiPerKur-PRL-00,GimGorSei-UNPUB-XX}. Moreover, by properly rescaling a given approximate functional $E_{\rm xc}[\rho]$, it is possible to test its performance in the strong-interaction limit, thus adding a new constraint for building approximations \cite{JunGarAlvGod-PRA-04,StaScuTaoPer-PRB-04,LiuBur-PRA-09}. More recently, the strong-interaction limit has been used to directly address strongly correlated systems \cite{LiuBur-JCP-09,GorSeiVig-PRL-09} in a DFT framework completely different with respect to the traditional KS one. 

In this work we aim at extending our knowledge on the functional $W_\alpha[\rho]$ by studying its behavior for $\alpha<0$, thus considering {\em attractive} electrons. In the very limit $\alpha\to-\infty$, we propose a rather simple and physically appealing solution, which can be evaluated exactly in the case of two-electron systems. The main interest in the exploration of this limit is to find new pieces of information on the unknown functional $W_\alpha[\rho]$. Similarly to previous work on the opposite $\alpha\to\infty$ limit, we expect that our results will open new possibilities for improving state-of-the art DFT approximations.
Notice that, so far, the $\alpha<0$ case has been only addressed in the special case of two electrons confined on the surface of a sphere, where accurate numerical calculations for $\alpha\in(-\infty,\infty)$ have been performed  \cite{Sei-PRA-07}. A functional tuned to reproduce these calculations also for $\alpha<0$ has been recently proposed \cite{Sun-JCTC-09}. The present work fully extends these early results by addressing the $\alpha\to-\infty$ limit in a general way.

The paper is organized as follows. In the next Sec.~\ref{sec_negativealpha} we present our solution for $\Psi_\alpha[\rho]$, $v_{\rm ext}^\alpha([\rho],\rv)$ and $W_\alpha[\rho]$ in the $\alpha\to-\infty$ limit, discussing the implications for DFT, as well as the limitations of our approach. 
As a first application, we show in Sec.~\ref{sec_EcGL2} that for two-electron systems, the information on the $\alpha\to-\infty$ limit can be used to get a rather accurate estimate of $E\c\glt[\rho]$ of Eq.~(\ref{EcGL2}) by simply interpolating between the $\alpha\to\pm\infty$ limits, and thus without using the unoccupied orbitals. The same procedure is also applied to the Be isoelectronic series, and the role of near-degeneracy effects in this framework is discussed. The last Sec.~\ref{sec_conc} is devoted to conclusions and perspectives.

\section{The limit $\alpha\to-\infty$}
\label{sec_negativealpha}

For $\alpha<0$, the Hamiltonian \eqref{Hv} describes {\em attractive} electrons
with a given ground-state density $\rho=\rho(\rv)$. To understand how
attractive fermions can be forced to form a given smooth density distribution
$\rho$ by means of a local external potential $v\ext^\alpha([\rho],\rv)$, we
consider here the extreme case $\alpha\to-\infty$.

\subsection{The wave function $\Psi_\alpha[\rho]$ for $\alpha\to-\infty$}
\label{Psineg}

As their attraction becomes very strong ($\alpha\to-\infty$), we expect that the
electrons in the state $\Psi_\alpha[\rho]$ form a compact
``attractive-electron cluster'' (AEC). With this, we mean that simultaneous measurement
of their positions in this state will always yield $N$ points $\rv_1,...,\rv_N$
in space that are very close to each other, much closer than any distance over
which the density $\rho(\rv)$ changes appreciably. For $N=2$, e.g.,
the AEC is a positronium-type object with an average distance of
$\langle|\rv_1-\rv_2|\rangle=\frac{2a_B}{|\alpha|}$ between the two electrons
(where $a_B=\hbar^2/m\e e^2$ is the Bohr radius). Notice that here we are only interested in the mathematical limit $\alpha\to-\infty$ of the Hamiltonian \eqref{Hv}, so that we disregard any relativistic effect.

\begin{figure}
\includegraphics[width=8.4cm]{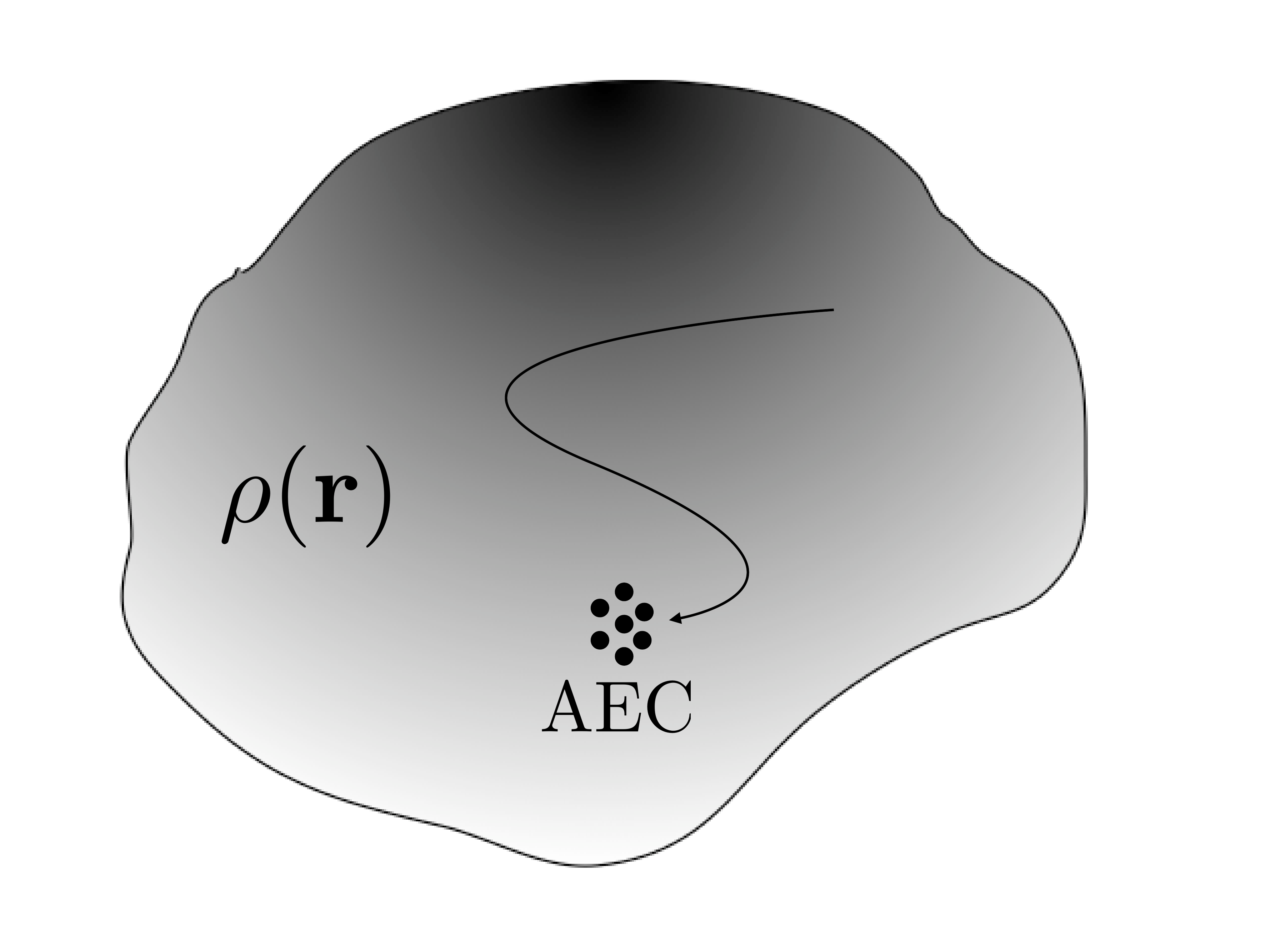} 
\caption{As the coupling strength $\alpha\to-\infty$, the $N$ electrons are expected to form a compact (point-like) attractive electron cluster (AEC), whose center of mass position has the probability distribution $\frac1N\rho(\rv)$.}
\label{fig_AEC}
\end{figure}

Exploiting this concept of a compact AEC, we 
expect that, as $\alpha\to-\infty$, the external potential $v\ext^\alpha([\rho],\rv)$
approaches a smooth function of $\rv$ which gives the quasi point-like AEC the probability
distribution $\frac1N\rho(\rv)$, as pictorially sketched in Fig.~\ref{fig_AEC}.
In order to formalize this idea, we write
\begin{eqnarray}
\sum_{i=1}^Nv\ext^\alpha([\rho],\rv_i)\,\Psi_\alpha([\rho];\rv_1,...,\rv_N;\sigma_1,...,\sigma_N)
\nonumber\\
\approx Nv\ext^\alpha([\rho],\rcm)\,\Psi_\alpha([\rho];\rv_1,...,\rv_N;\sigma_1,...,\sigma_N)
\nonumber\\
\qquad(\alpha\ll-1),
\label{vextapp}
\end{eqnarray}
where $\rcm=\frac1N\sum_{i=1}^N\rv_i$ is the center of mass of the $N$ electrons.
The accuracy of this approximation grows indefinitely as
$\alpha\to-\infty$, when the radius of the AEC tends to zero.
Thus, we introduce relative Jacobi coordinates $\sv_n$ ($n=1,...,N-1$)
and the center-of-mass $\sv_N\equiv\rcm$,
\begin{eqnarray}
\sv_n&=&\frac1n\sum_{i=1}^{n}\rv_i-\rv_{n+1}\qquad(1\le n<N),\label{coos}\\
\sv_N&=&\frac1N\sum_{i=1}^N\rv_i\;\equiv\;\rcm.\label{coosN}
\end{eqnarray}
For $1<n<N$, the inverse transformation reads
\beq
\rv_n=
\rcm-\frac{n-1}n\,\sv_{n-1}
    +\sum_{\ell=n}^{N-1}\frac{\sv_{\ell}}{\ell+1},
\label{inverse}
\eeq
while for $n=1$ and $n=N$ we have
\beq
\rv_1=\rcm+\sum_{\ell=1}^{N-1}\frac{\sv_{\ell}}{\ell+1},\qquad
\rv_N=\rcm-\frac{N-1}N\,\sv_{N-1}.
\label{inverse1N}
\eeq

In terms of the Jacobi coordinates $\sv_1,...,\sv_{N-1},\sv_N\equiv\rcm$, the operator
$\hat{T}=-\frac{\hbar^2}{2m\e}\sum_{i=1}^{N}\nabla_i^2$ assumes the form
\beq
\hat{T}=-\frac{\hbar^2}{2M}\frac{\partial^2}{\partial\rcm^2}+\hat{T}\rel~,
\qquad M=Nm\e,
\label{Trel}
\eeq
where $\hat{T}\rel$ acts on the relative coordinates only,
\beq
\hat{T}\rel=-\frac{\hbar^2}2
\sum_{n=1}^{N-1}\frac1{m_n}\frac{\partial^2}{\partial\sv_n^2},\qquad
m_n=\frac{n}{n+1}m\e.
\label{TrelE}
\eeq
Also, the purely multiplicative operator $\hat{V}\ee$ of Eq.~\eqref{Vee}
depends on the relative coordinates only,
\beq
\hat{V}\ee=\widetilde{V}\ee(\sv_1,...,\sv_{N-1}),
\eeq
since $\rv_i-\rv_j$ is independent of $\rcm$, see Eqs.~\eqref{inverse} and \eqref{inverse1N}.

The resulting structure of the Hamiltonian \eqref{Hv}, within the approximation
\eqref{vextapp}, implies a product ansatz for the wave function
$\Psi_\alpha[\rho]$ in terms of the new coordinates,
\begin{eqnarray}
\Psi_\alpha[\rho]\;\to\;
\phi_\alpha([\rho];\rcm)\;\psi_\alpha(\sv_1,...,\sv_{N\!-\!1};\sigma_1,...,\sigma_N)
\label{product}
\end{eqnarray}
where $\phi_\alpha$ and $\psi_\alpha$ are, respectively, the
lowest-eigenvalue solutions of the following Schr\"odinger Equations,
\begin{eqnarray}
\Big\{-\frac{\hbar^2}{2M}\frac{\partial^2}{\partial\rcm^2}
+Nv\ext^\alpha([\rho],\rcm)\Big\}\phi_\alpha
&=&E\cm^\alpha\phi_\alpha~,\qquad\label{SEqphi}\\
\Big\{\hat{T}\rel+\alpha\widetilde{V}\ee(\sv_1,...,\sv_{N\!-\!1})\Big\}\psi_\alpha
&=&E\rel^\alpha\psi_\alpha~.\qquad\label{SEqpsi}
\end{eqnarray}
Since $\phi_\alpha([\rho];\rcm)$ is symmetric with respect to permutations of
the electronic coordinates $\rv_n$, the second factor $\psi_\alpha$ of the 
wave function \eqref{product} must be anti-symmetric,
\beq
\psi_\alpha(...)
=\frac1{\sqrt{N!}}\sum_{\pi\in S_N}(-1)^\pi\hat{P}_\pi\psi_\alpha(...).
\eeq
Here, $S_N$ is the group of the $N!$ permutations $\pi$ of $N$ elements,
and $(-1)^\pi$ is the sign of $\pi$. The operator $\hat{P}_\pi$ is defined by
\begin{eqnarray}
\hat{P}_\pi\psi_\alpha(\{\sv(\rv_1,...,\rv_N)\};\sigma_1,...,\sigma_N)
=\hspace*{1.25cm}
\nonumber\\
\psi_\alpha\big(\{\sv(\rv_{\pi(1)},...,\rv_{\pi(N)})\};
                  \sigma_{\pi(1)},...,\sigma_{\pi(N)}\big),
\label{Ppi}
\end{eqnarray}
with the short-hand notation $\{\sv\}:=\sv_1,...,\sv_{N-1}$ and the explicit
functions $\sv_n=\sv_n(\rv_1,...,\rv_N)$ of Eq.~\eqref{coos}.
To recover on the right-hand side a pure function of $\sv_1,\dots,\sv_{N-1}$ (and of the $\sigma_n$),
the $\rv_{\pi(n)}$ must be re-expressed in terms of the $\sv_n$
by Eqs.~\eqref{inverse} and \eqref{inverse1N}, where $\sv_N$ cancels out.

The factors $\phi_\alpha$ and $\psi_\alpha$ of the wave function \eqref{product}
are addressed, respectively, in the following two subsections.

\subsection{The external potential $v\ext^\alpha([\rho],\rv)$ for
$\alpha\to-\infty$}

In terms of the first factor $\phi_\alpha$ in Eq.~\eqref{product},
the probability density of the electronic center-of-mass position $\rcm$ in the
wave function $\Psi_\alpha[\rho]$ for $\alpha\to-\infty$ is given by
\begin{eqnarray}
\rho\cm^\alpha([\rho],\rcm)&\to&\!\!\!\sum_{\sigma_1,...,\sigma_N}
\int\!\!d^3s_1\cdots\int\!\!d^3s_{N\!-\!1}\;  |\phi_\alpha\psi_\alpha|^2\nonumber\\
&=&|\phi_\alpha([\rho],\rcm)|^2.
\end{eqnarray}
For $\alpha\to-\infty$, when the AEC becomes very compact (point-like), the function
$N\rho\cm^\alpha([\rho],\rcm)$ must approach the electron density $\rho(\rv)$.
Then, using $\phi_\alpha(\rv)=\sqrt{\rho\cm^\alpha(\rv)}$, Eq.~\eqref{SEqphi}
can be resolved for $v\ext^\alpha([\rho],\rv)$,
\beq
\lim_{\alpha\to-\infty}v\ext^\alpha([\rho],\rv)=\frac{E\cm^{-\infty}}N
+\frac{\hbar^2}{2m\e N^2}\frac{\nabla^2\sqrt{\rho(\rv)}}{\sqrt{\rho(\rv)}}.
\label{limpot}
\eeq
According to this result, the external potential $v\ext^\alpha([\rho],\rv)$ in
the Hamiltonian \eqref{Hv} approaches a smooth and finite function
$v\ext^{-\infty}([\rho],\rv)$, as $\alpha\to-\infty$. The value $E\cm^{-\infty}$
can be fixed by the condition $v\ext^\alpha([\rho],\rv)\to0$ for $|\rv|\to\infty$.
Obviously, $Nv\ext^{-\infty}([\rho],\rv)$ is simply the external potential for a
single particle with mass $M=Nm\e$ (which is the AEC of the previous Subsec.~\ref{Psineg}) and
ground-state density $\frac1N\rho(\rv)$.

So far, we see that an external potential $v\ext^\alpha([\rho],\rv)$ can,
for $\alpha\to-\infty$, generate a given smooth density distribution $\rho(\rv)$
for $N$ attractive electrons.
The smooth distribution is achieved by the uncertainty in the center-of-mass
position $\rcm$ of the AEC. This picture 
becomes questionable in cases such as the stretched H$_2$ molecule, where
the density is separated spatially into several pieces.

\subsection{The integrand $W_\alpha[\rho]$ for $\alpha\to-\infty$}
As we have seen in Subsec~\ref{Psineg}, as $\alpha\to-\infty$ $\Psi_\alpha[\rho]$ should be better and better approximated by the product ansatz of Eq.~\eqref{product}. Consequently, the expectation $\langle\Psi_\alpha[\rho]|\hat{V}\ee|\Psi_\alpha[\rho]\rangle$ can be computed in this limit by using the second factor $\psi_\alpha$ of Eq.~\eqref{product}, as $\hat{V}_{ee}$ only depends on the relative coordinates. Thus, for $\alpha\to-\infty$ we have
\begin{eqnarray}
W_\alpha[\rho]+U[\rho]\equiv\langle\Psi_\alpha[\rho]|\hat{V}\ee|\Psi_\alpha[\rho]\rangle\;\to
\hspace{2cm}
\nonumber\\
\sum_{\sigma_1,...,\sigma_N}
\int\!\!d^3s_1\cdots\int\!\!d^3s_{N\!-\!1}
|\psi_\alpha(...)|^2\widetilde{V}\ee(\sv_1,...,\sv_{N\!-\!1}), 
\label{psiVeepsi}
\end{eqnarray}
where we have integrated out the center of mass coordinate and used the normalization $\int d^3R|\phi_\alpha([\rho],\rcm)|^2=1$. We now remark that the Schr\"odinger equation for $\psi_\alpha$  of Eq.~\eqref{SEqpsi} is not affected by the details of the density function $\rho(\rv)$, but only by
the electron number $N=\int d^3r\rho(\rv)$, so that the same is true for its lowest-eigenvalue
solution $\psi_\alpha$ and thus for the right-hand side of Eq.~\eqref{psiVeepsi}. This
means that, for all different densities $\rho$ with the same electron number $N$,
the function $W_\alpha[\rho]+U[\rho]$ on the left-hand side of Eq.~\eqref{psiVeepsi} has the same asymptote when $\alpha\to-\infty$. This point is of course very appealing, but has obvious limitations that will be discussed in the next Subsec.~\ref{sec_limitations}. 

To address the solution $\psi_\alpha$ of Eq.~\eqref{SEqpsi}, we consider the
universal (i.e., density-independent) hamiltonian
\beq
\hat{H}_\alpha^N=\hat{T}+\alpha\hat{V}\ee\qquad(\alpha<0).
\label{Huniv}
\eeq
$\hat{H}_\alpha^N$ describes a system of $N$ {\em attractive} electrons in the
{\em absence} of any localizing external potential $v\ext(\rv)$. Such a system has
translational symmetry and a uniform ground-state density $\bar{\rho}$. To make
$\bar{\rho}$ finite, we can introduce a normalization volume $\itOm=N/\bar{\rho}$
and impose periodic boundary conditions on the wave function. In the ground state
of $\hat{H}_\alpha^N$, the $N$ electrons are forming a ``free'' AEC whose
center-of-mass position $\rcm$ has a uniform probability distribution within
$\itOm$, with density $\frac1N\bar{\rho}$.

Notice that for the hamiltonian \eqref{Huniv}, Eq.~\eqref{vextapp} is not an approximation, since $v\ext^\alpha([\bar{\rho}],\rv)\equiv0$.
Consequently, the product on the right-hand side of Eq.~\eqref{product}, where the factors
$\phi_\alpha$ and $\psi_\alpha$ satisfy Eqs.~\eqref{SEqphi}
[with $v\ext^\alpha\equiv0$] and \eqref{SEqpsi},
respectively, represents the exact ground state $\Psi_\alpha^N$ of $\hat{H}_\alpha^N$,
for {\em all finite} values of $\alpha\leq0$. Therefore, Eq.~\eqref{psiVeepsi} can be
written as
\beq
\langle\Psi_\alpha[\rho]|\hat{V}\ee|\Psi_\alpha[\rho]\rangle\to
\langle\Psi_\alpha^N|\hat{V}\ee|\Psi_\alpha^N\rangle\quad(\alpha\to-\infty).
\label{asymp}
\eeq



Due to the universal form of the Hamiltonian \eqref{Huniv}, its ground-state
wave functions $\Psi_\alpha^N$ for different interaction strengths $\alpha<0$
are related by a simple scaling law,
\beq
\Psi_\alpha^N\big(\{\rv,\sigma\}\big)
=|\alpha|^{3N/2}\Psi_{\alpha=-1}^N\big(\{|\alpha|\rv,\sigma\}\big),
\label{scaPsi}
\eeq
in the short-hand notation
$\{\rv,\sigma\}\equiv\rv_1,...,\rv_N;\sigma_1,...,\sigma_N$.
The resulting scaling law for the ground-state energy $E_\alpha^N$ of
$\hat{H}_\alpha^N$ reads
\beq
E_\alpha^N=\alpha^2E_{\alpha=-1}^N.
\label{scalE}
\eeq
Notice that $E_\alpha^N$ is the eigenvalue in Eq.~\eqref{SEqpsi},
\beq
\Big\{\hat{T}\rel+\alpha\widetilde{V}\ee(\sv_1,...,\sv_{N\!-\!1})\Big\}\psi_\alpha
=E^N_\alpha\psi_\alpha~.\qquad\label{SEqpsi2}
\eeq
The virial theorem, $\langle\Psi_\alpha^N|\hat{T}|\Psi_\alpha^N\rangle=
-\frac12\langle\Psi_\alpha^N|\alpha\hat{V}\ee|\Psi_\alpha^N\rangle$, yields
$E_\alpha^N=\frac12\langle\Psi_\alpha^N|\alpha\hat{V}\ee|\Psi_\alpha^N\rangle$
or
\beq
\langle\Psi_\alpha^N|\hat{V}\ee|\Psi_\alpha^N\rangle=\frac2{\alpha}E_\alpha^N=2\alpha\,E^N_{-1}.
\eeq
Consequently, according to Eq.~\eqref{asymp}, the integrand of Eq.~\eqref{CCI}
asymptotically approaches for $\alpha\to-\infty$ a linear function with slope
$2E_{-1}^N<0$,
\beq
W_\alpha[\rho]\;\to\;2\alpha\,E_{-1}^N-U[\rho]\qquad(\alpha\to-\infty).
\label{sal}
\eeq
As said, since $E_{-1}^N$ is the ground-state energy of a free AEC at $\alpha=-1$, it cannot
depend on details of the density $\rho$, but only on the electron number
$N=\int d^3r\,\rho({\bf r})$. We discuss below the limitations of this result. 

\subsection{Limitations of the AEC solution}
\label{sec_limitations}
First of all, as already mentioned, the AEC picture breaks down when the density is composed of several disjointed subsystems (e.g., stretched molecules). In the case of non-interacting fragments, in order to be size consistent, the limit of Eq.~\eqref{psiVeepsi} should apply separately to each subsystem, which is physically reasonable, but difficult to prove rigorously. The transition from a jointed to a disjointed system (e.g., during molecular stretching) remains also unclear.

An even more important point is the following. In order to be useful to construct approximations for $W_\alpha[\rho]$, the $\alpha\to-\infty$ 
asymptote of Eq.~\eqref{psiVeepsi} should be approached in a continuous and smooth way starting from $\alpha=1$ (i.e., without phase transitions). This is generally not true, so that often the state $\psi_\alpha$ which is smoothly connected with the physical ($\alpha=1$) system is not the ground state of the hamiltonian of Eq.~\eqref{SEqpsi}. As a simple example, consider the N atom. At $\alpha=1$, we have $N_\uparrow=5$ spin-up electrons and $N_\downarrow=2$ spin-down electrons, but we may expect that the ground state of the attractive electron cluster corresponds to $N_\uparrow=4$, $N_\downarrow=3$, as pairing is usually energetically advantageous for an attractive interaction, so that we will have a spin-flip transition at $\alpha<0$. This simple case might be solved in a spin-DFT framework (in which we look for the $\alpha<0$ systems with the same spin densities of the physical system), but we may encounter much more difficult cases as the number of electrons increases, with possible ``exotic'' phases. An interesting point is that, however, the $\alpha\to-\infty$ potential $v\ext^\alpha([\rho],\rv)$ of Eq.~\eqref{limpot} is the same for all the excited states of the cluster, including the one continuously connected with the physical system, if such a state exists. 
Nonetheless, with all these limitations in mind, we may still expect that the results presented in the previous section be qualitatively correct for many interesting systems.

In simple cases such as closed-shell two- and four-electron systems with a reasonable compact density, we might expect a continuous connection between the limit of Eq.~\eqref{sal} and the $\alpha=1$ case. In the next Sec.~\ref{sec_EcGL2}, we construct a simple continuous interpolation between the two limits $\alpha\to\pm\infty$  and we use it to estimate $E\c\glt[\rho]$ for spherical two- and four-electron densities. In the latter case, we will discuss the role of near-degeneracy effects. 

\section{$E\c\glt[\rho]$ from the limits $\alpha\to\pm\infty$}
\label{sec_EcGL2}
We now construct an analytic model $W\mdl_{\alpha}[\rho]$ (``AR'' stands for
``attraction-repulsion'' here) that shares the properties
\eqref{wil}, \eqref{sil}, and \eqref{sal} with the unknown exact integrand 
$W_{\alpha}[\rho]$ of Eq.~\eqref{CCI}, which we assume here to be a continuous and smooth
function of $\alpha$ also for $\alpha\leq0$, even if we keep in mind the limitations of this assumption discussed in the previous Subsec.~\ref{sec_limitations}. In particular, our model will share
the correct density-scaling behavior with the unknown exact integrand $W_{\alpha}[\rho]$,
see Eq.~\eqref{Wscal} below. Since the {\em inverse} function is easier to model, we write
\beq
W\mdl_{\alpha}[\rho]\;=\;f^{-1}(\alpha).
\label{fm1}
\eeq
According to Eqs.~\eqref{wil}, \eqref{sil}, and \eqref{sal}, respectively,
this inverse function $f(\xw)$ must satisfy the conditions
\begin{eqnarray}
f(\xw) & \to & \left(\frac{W'_{\infty}}{\xw-W_{\infty}}\right)^2\hspace*{1.0cm}(\xw\to W_{\infty}),
\label{grequ1}\\
f(E\x)\hspace*{-0.15cm} &  =  & 0\hspace*{3.3cm}(w=E\x),
\label{grequ2}\\
f(\xw)   & \to & \frac{\xw+U}{2E_{-1}^N}\hspace*{2.4cm}(\xw\to\infty ).
\label{grequ3}
\end{eqnarray}
Here, we have dropped the functional symbol $[\rho]$ for brevity, as we shall do
in most of the following equations.

\begin{figure}
\includegraphics[width=8.4cm]{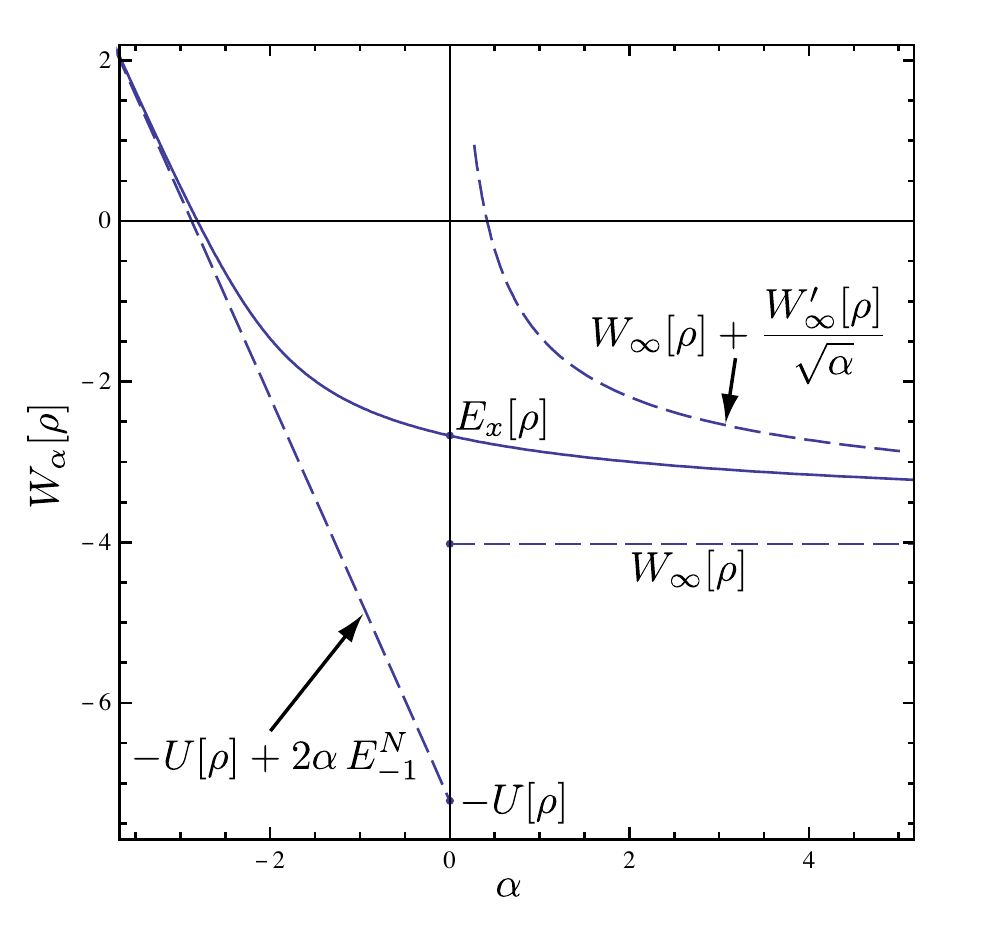} 
\caption{The functional $W_\alpha[\rho]$ as a function of $\alpha$ (solid curve), as modeled in Sec.~\ref{sec_EcGL2}, using the five parameters for the Be atom in Table \ref{tab_EcGL2}.}
\label{fig_Walphamod}
\end{figure}

\begin{figure}
\includegraphics[width=8.4cm]{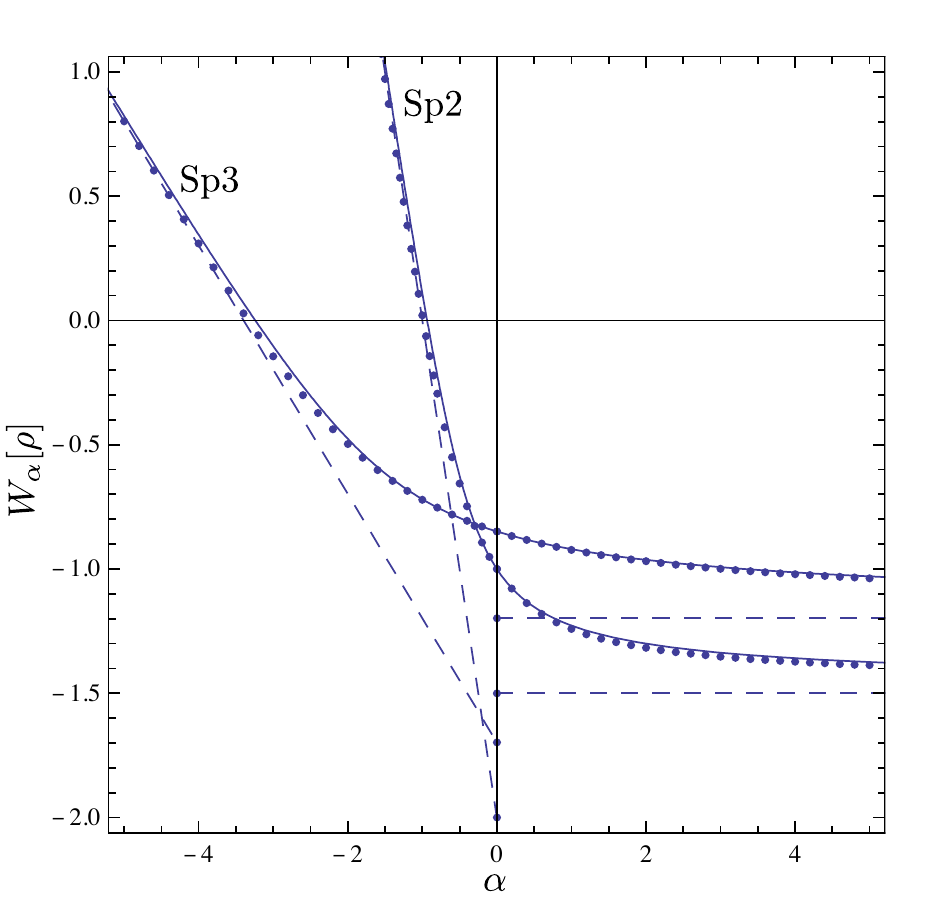} 
\caption{The functional $W_\alpha[\rho]$ as a function of $\alpha$ for the systems `Sp2' and `Sp3' (see Appendix \ref{appElSp}). The accurate numerical values (dots) are compared with the model of Sec.~\ref{sec_EcGL2} (solid curves).}
\label{fig_Sp2Sp3}
\end{figure}

Eqs.~\eqref{grequ1} and \eqref{grequ3}
are satisfied by the preliminary choice 
\beq
f_1(\xw)\;=\;\frac{\xw+U}{2E_{-1}^N}
       +\left(\frac{W'_{\infty}}{\xw-W_{\infty}}\right)^2
\label{f1}
\eeq
which, however, ignores condition \eqref{grequ2}.
In particular, $f_1(\xw)$ approaches the asymptotic form of Eq.~\eqref{grequ3} too slowly,
$f_1(w)=(w+U)/2E^N_{-1}+O(\xw^{-2})$. 
Instead, we expect $f(w)=(w+U)/2E^N_{-1}+O(e^{-w})$ or, equivalently,
\beq
W_\alpha[\rho]=2\alpha\,E_{-1}^N-U[\rho]+O(e^{\alpha}) \qquad (\alpha\to-\infty).
\eeq
This conjecture is based on Eq.~\eqref{scaPsi}, indicating that the size of the
AEC shrinks by the factor $|\alpha|^{-1}$ as $\alpha\to-\infty$, while the
external potential $v\ext^{\alpha}([\rho],\rv)$ approaches a smooth finite limit,
see Eq.~\eqref{limpot}. Consequently, the point where the radius of the AEC becomes
smaller than any distance $|\rv|$ over which $v\ext^{\alpha}([\rho],\rv)$ changes
appreciably, must be reached at some moderately negative value $\alpha_0[\rho]$ of
the parameter $\alpha$,
\beq
W_\alpha[\rho]\;\approx\;2\alpha\,E_{-1}^N-U[\rho]\qquad(\alpha<\alpha_0[\rho]).
\label{alph0}
\eeq
This conjecture is also supported by the numerical data sets shown in Fig.~\ref{fig_Sp2Sp3},
where $\alpha_0$ takes values $\approx-3.4$ (`Sp3') and $\approx-1.0$ (`Sp2').
Accordingly, we modify Eq.~\eqref{f1},
\beq
f_2(\xw)\;=\;\frac{\xw+U}{2E_{-1}^N}+\left( \frac{W'_{\infty}}{\xw-W_{\infty}}\right)^2
e^{-(\xw-W_{\infty})/\itGa}.     
\label{f2}
\eeq
Note that the exponential factor in Eq.~\eqref{f2} equals 1 at $w=W_{\infty}[\rho]$,
thus preserving the correct limit \eqref{grequ1}.
Due to the general scaling property of the exact $W_\alpha[\rho]$, see Eq.~\eqref{Wscal}
below, the parameter $\itGa=\itGa[\rho]$ must scale the same way as $W_{\infty}[\rho]$,
$\itGa[\rho_\lambda]=\lambda\itGa[\rho]$. Therefore, we set
\beq
\itGa[\rho]=|W_{\infty}[\rho]|,
\eeq
identifying the unknown number $\alpha_0[\rho]$ from Eq.~\eqref{alph0} approximately
with the zero of $W_{\alpha}[\rho]$,
\beq
W_{\alpha_0[\rho]}[\rho]\;\approx\;0\qquad\Leftrightarrow\qquad f(0)\;\approx\;\alpha_0[\rho],
\label{alpha0}
\eeq
see Figs.~\ref{fig_Walphamod} and \ref{fig_Sp2Sp3}.

To satisfy condition \eqref{grequ2}, too, we introduce an extra factor $h_B(\xw)$
with $h_B(W_{\infty})=1$,
\beq
f(\xw)\;=\;\frac{\xw+U}{2E_{-1}^N}+\left( \frac{W'_{\infty}}{\xw-W_{\infty}}\right)^2
h_B(\xw)\; e^{-(\xw-W_{\infty})/\itGa}.        
\label{f}
\eeq
In terms of the number
\beq
B\;\equiv\;\frac{E\x +U}{-2E_{-1}^N}\left(\frac{E\x -W_{\infty}}{W'_{\infty}}\right)^2
e^{(E\x -W_{\infty})/\itGa}\;>\;0,             
\label{B}
\eeq
we choose for $\xw\in[W_{\infty},+\infty]$
\beq
h_B (\xw) =
\left\{\begin{array}{l@{\quad:\quad}l}
\;\,  1+           (B-1)      \,\frac{\xw-W_{\infty}}{E\x-W_{\infty}} & B\geq 1, \\
\left[1+\left(\frac1B-1\right)\!\frac{\xw-W_{\infty}}{E\x-W_{\infty}}\right]^{-1}
                                                                      & B\leq 1.
         \end{array}\right.
\label{hB}
\eeq
Then, $h_B(W_{\infty})=1$, $h_B(E\x)=B$, and $h_B(\xw)>0$ is monotonically
increasing ($B>1$) or decreasing ($B<1$).

The function $W_{\alpha}\mdl[\rho]=f^{-1}(\alpha)$ has five parameters:
$E_{-1}^N$, $U[\rho ]$, $E\x[\rho ]$, $W_{\infty}[\rho]$, and $W'_{\infty}[\rho]$, and it is shown
in Fig.~\ref{fig_Walphamod}. As an illustration, Fig.~\ref{fig_Sp2Sp3} shows the true integrand $W_\alpha[\rho]$ (dots), evaluated numerically for the systems `Sp2' and `Sp3' (see Appendix \ref{appElSp}). For these systems, the evaluation of $W_\alpha[\rho]$ is straightforward, since their $\alpha$-dependent external potential $v_{\rm ext}^{(\alpha)}(\rv)$ in the Hamiltonian \eqref{Hv} is trivial and thus known from the beginning.

Since $2E\c\glt[\rho]\equiv \frac{d}{d\alpha}W_{\alpha}[\rho]|_{\alpha=0}
\approx1/f'(E\x)$, our model $W_\alpha^{\rm AR}[\rho]$ yields a simple prediction
$\widetilde{E}\c\glt[\rho]\equiv\frac1{2f'(E\x)}$ for the second-order correlation energy
 $E\c\glt[\rho]$,
\begin{eqnarray}
\widetilde{E}\c\glt[\rho ]=
E_{-1}^N\left[1+(E\x+U)\left(\displaystyle\frac{B'}{E\x-W_{\infty}}-\frac1{W_{\infty}}\right)\right]^{-1}
\nonumber\\
B'\;=\;\left\{\begin{array}{l}
1+\frac1B\quad(B\geq1),\\3-B\quad(B\leq1).\end{array}\right.\hspace*{0.4cm}
\label{Ec2MOD}
\end{eqnarray}
Unlike the exact functional $E\c\glt[\rho ]$ whose evaluation requires additional knowledge
of all the unoccupied KS orbitals, the approximation \eqref{Ec2MOD} only depends on the $N$
occupied orbitals (via $E\x[\rho]$), on the universal number $E_{-1}^N$,
and explicitly on the density $\rho$ itself.
Nevertheless, the new functional has the correct scaling behavior of the exact $E\c\glt[\rho ]$ (only valid when the KS system does not become degenerate or quasi-degenerate),
\beq
\widetilde{E}\c\glt[\rho_\lambda]\;=\;\widetilde{E}\c\glt[\rho],
\label{eq_scalingEcGL2}
\eeq
where $\rho_\lambda({\bf r})\equiv\lambda^3\rho(\lambda{\bf r})$ (with $\lambda>0$) is a
scaled density.
More generally, the model integrand $W_{\alpha}\mdl[\rho]$ has the correct
scaling property \cite{Lev-INC-87} of its exact counterpart $W_{\alpha}[\rho]$,
\beq
W_{\alpha}\mdl[\rho_\lambda]\;=\;\lambda\,W_{\alpha/\lambda}\mdl[\rho].
\label{Wscal}
\eeq
This law has a simple graphical interpretation: Plotting $W_{\alpha}\mdl[\rho_\lambda]$
versus $\alpha$ amounts to zooming the corresponding plot of $W_{\alpha}\mdl[\rho]$ by
the factor $\lambda$. This graphical property is satisfied by the function $f(\xw)$,
Eq.~\eqref{f} (and then also for its inverse, $W_{\alpha}\mdl[\rho]$), since we have
the general scaling laws
$U[\rho_\lambda]\!=\!\lambda\,U[\rho]$,
$E\x[\rho_\lambda]\!=\!\lambda\,E\x[\rho]$,
$W_{\infty}[\rho_\lambda]\!=\!\lambda\,W_{\infty}[\rho]$, and
$W'_{\infty}[\rho_\lambda]\!=\!\lambda^{3/2}\,W'_{\infty}[\rho]$.

\begin{table*}
\begin{tabular}{|l|r|r||rrrrr||c||rr|c|}
\hline
System  & $N$ & $D$ & $E_{-1}^N$ & $U[\rho]$ & $E\x[\rho]$ &
$W_{\infty}[\rho]$ & $W'_{\infty}[\rho]$ &
$B$ & $\widetilde{E}\c\glt[\rho]$ & $E\c\glt[\rho]$ & error \\ \hline
Units & - & - & \multicolumn{5}{|c||}{1 Ha=27.21 eV} & - & \multicolumn{2}{|c|}{$10^{-3}$ Ha} & $10^{-3}$ Ha \\ \hline
`Sp2'  & 2  & 2 & $-$1.00
                     & 2.000 & $-$1.000 & $-$1.500 & 0.250 & 2.791 & $-$228.1 & $-$227.4 & $-$0.7 (0.3\%) \\
`Sp3'  & 2  & 3 & $-$0.25 &  1.698 & $-$0.849 & $-$1.198 & 0.375 & 1.968 & $-$46.5 & $-$47.6 & $+$1.1 (2.3\%) \\
`Exp'  & 2  & 3 & $-$0.25 &  1.250 & $-$0.625 & $-$0.910 & 0.345 & 1.167 & $-$43.4 & $-$46.7 & +3.3 (7.1\%) \\
`Hoo'  & 2  & 3 & $-$0.25 &  1.030 & $-$0.515 & $-$0.743 & 0.208 & 1.682 & $-$47.2 & $-$50.5 & +3.3 (6.5\%) \\ 
 He    & 2 & 3 & $-$0.25 &  2.049 & $-$1.025 & $-$1.500 & 0.621 & 1.649 & $-$48.6 & $-$47.5 & $-$1.1 (2.3\%) \\
Ne$^{8+}$ &2 & 3 & $-$0.25 & 12.055 & $-$6.028 & $-$8.794 & 8.792 & 1.631 & $-$48.0 & $-$46.7 & $-$1.3 (2.8\%) \\
Be  & 4 & 3 & $-$1.255 & 7.217 & $-$2.673 & $-$4.021 & 2.59 & 0.6857 & $-$126.4 & $-$128.4 & $+$2.0 (1.6\%) \\
Ne$^{6+}$  & 4 & 3  & $-$1.255 & 21.742 & $-$7.600 & $-$11.563 & 12 & 0.8655 & $-$127.5 & $-$320.4 & $+$193 (60\%) \\
\hline
\end{tabular}
\caption{
The AEC energies $E_{-1}^N$ and the coefficients $U[\rho ]$, $E\x[\rho ]$,
$W_{\infty}[\rho ]$, and $W'_{\infty}[\rho ]$ of the expansions \protect\eqref{sal},
\protect\eqref{wil}, and \protect\eqref{sil},
the dimensionless parameter $B$, Eq.~\protect\eqref{B}, and
the resulting predictions $\widetilde{E}\c\glt[\rho ]$ of the second-order correlation
energy $E\c\glt[\rho ]$ for various $D$-dimensional $N$-electron systems.
The systems `Sp2' and `Sp3', here with radius $R=a_B$, are defined in
Appendix \ref{appElSp}; `Exp' refers to a hypothetic
two-electron atom with ground state density $\rho(r)=e^{-r/a_B}/4\pi a_B^3$;
`Hoo' refers to the Hooke atom, consisting of two electrons in the external potential
$v\ext(\rv)=\frac{k}{2}r^2$, with $k=\frac{1}{4}$.
}
\label{tab_EcGL2}
\end{table*}

\subsection{Application to $N=2$ systems}
\label{subsec_EcGL2N2}
For a closed-shell two-electron system with a reasonable compact density, the cluster energy $E_{-1}^{N=2}$ needed in Eq.~\eqref{sal} can be calculated exactly: the AEC defines a positronium-like problem whose ground state energy is $E_{-1}^{2}=\frac14$ Ha. Moreover, in this special case we expect that the $\alpha\to-\infty$ limit is reached smoothly from $\alpha=1$. 

In Table~\ref{tab_EcGL2} we report the prediction of $E\c\glt[\rho ]$ from Eq.~\eqref{Ec2MOD} for several two-electron densities, and we compare it with accurate values from the literature. The systems denoted  `Sp2' and `Sp3' correspond to electrons confined on the surface of a sphere (in 2D and 3D, respectively) and are defined in Appendix \ref{appElSp}. For these systems exact values are available~\cite{Sei-PRA-07,LooGil-PRA-09,LooGil-PRL-09}. `Exp' refers to a hypothetic two-electron atom with ground state density $\rho(r)=e^{-r/a_B}/4\pi a_B^3$, whose corresponding accurate $E\c\glt[\rho ]$ is taken from Ref.~\onlinecite{IvaLev-JPCA-98}. `Hoo' refers to the Hooke atom, consisting of two electrons in the external potential $v\ext(\rv)=\frac{k}{2}r^2$, with $k=\frac{1}{4}$. Its exact density is analytically known \cite{Tau-PRA-93}, and the corresponding $E\c\glt[\rho ]$ is from Ref.~\onlinecite{MagTerBur-JCP-03}. The accurate densities for He and Ne$^{8+}$ and the corresponding values of $E\c\glt[\rho ]$ are taken from the work of Colonna and Savin \cite{ColSav-JCP-99}. For all the densities considered here, the functionals $W_\infty[\rho]$ and $W_\infty'[\rho]$ have been evaluated using the procedure described, respectively, in Refs.~\onlinecite{SeiGorSav-PRA-07} and \onlinecite{GorVigSei-JCTC-09}. 

We see that for all the two-electron systems considered here, the functional $\widetilde{E}\c\glt[\rho ]$ of Eq.~\eqref{Ec2MOD} is in very good agreement with the corresponding accurate values from the literature, with a maximum error of $\sim 3$ mHa. Notice in particular how, for the `Sp2' system, the unusually high value $E_c^{\rm GL2}[\rho]=-227.4$ mHa of the slope $\frac12\frac{d}{d\alpha}W_\alpha[\rho]|_{\alpha=0}$ in Fig.~\ref{fig_Sp2Sp3} is predicted from the unusually high energy $E_{-1}^2=-1$ Ha (i.e., the slope of the corresponding inclined dashed line at $\alpha<0$ in Fig.~\ref{fig_Sp2Sp3}) of the AEC in two dimensions.  In contrast, the remaining four parameters $U$, $E_x$, $W_\infty$, and $W'_\infty$ in Table \ref{tab_EcGL2} do not differ dramatically between `Sp2' and `Sp3'.

\subsection{Application to the Be series: \\ The effect of near-degeneracy}
For the Be isoelectronic series, we have closed-shell $N=4$ systems, so that we may still expect that the $\alpha\to-\infty$ limit of Eq.~\eqref{sal} is approached smoothly from $\alpha=1$. However, as the nuclear charge $Z e$ increases, the 2$s$ and 2$p$ KS orbital energies become closer and closer, resulting in a large increase of $E\c\glt[\rho ]$. Indeed, in this case the scaling property of Eq.~\eqref{eq_scalingEcGL2}, satisfied by the present model of Eq.~\eqref{Ec2MOD}, does not hold anymore, so that we can expect a failure of Eq.~\eqref{Ec2MOD} for large $Ze$. 

As an illustration, we have applied the model of Eq.~\eqref{Ec2MOD} to the Be and the Ne$^{6+}$ densities. The cluster energy for $N=4$ has been estimated from a configuration interaction (CI) calculation \cite{Pin-THESIS-00}. The accurate values of $U[\rho]$, $E_x[\rho]$ and $E\c\glt[\rho ]$  are taken, in both cases, from the work of Colonna and Savin~\cite{ColSav-JCP-99}. Using the same densities of Colonna and Savin \cite{ColSav-JCP-99}, we have calculated here the values of $W_\infty[\rho]$ and $W_\infty'[\rho]$, following the procedures described, respectively, in Refs.~\onlinecite{SeiGorSav-PRA-07} and \onlinecite{GorVigSei-JCTC-09}.

We see that the result for Be is quite accurate, yielding an estimate of  $E\c\glt[\rho ]$ with an error of only 2 mH, confirming the hypothesis that, also in this case, the $\alpha\to-\infty$ limit is reached smoothly from $\alpha=1$. However, as the system starts to display strong near-degeneracy as in the case of Ne$^{6+}$, we see that the present model fails, making an error of 60\% in the estimate of  
$E\c\glt[\rho ]$.
When $Ze$ continues to increase beyond the value $Z=10$ considered here, the ground-state density becomes no longer pure-state $v$-representable, but only ensemble $v$-representable \cite{Lev-PRA-82,Lie-IJQC-83,ChaChaRus-JSP-85,SchGriBae-TCA-98,SchGriBae-JCP-99,UllKoh-PRL-01}. At this point, the definition of both $E_x[\rho]$ and $E\c\glt[\rho ]$ should change.

\section{Conclusions and perspectives}
\label{sec_conc}
We have presented a comprehensive analysis of the DFT adiabatic connection at negative coupling strengths $\alpha$. In the extreme limit $\alpha\to-\infty$ we have found a simple and physically appealing solution, which, albeit suffering from limitations due to the complexity of the many-electron problem with attractive interaction, can be calculated exactly in the case of reasonably compact two-electron systems, and provides insight also when $N>2$. As a first example of application of our results, we have shown that, for $N=2$, the exact information at $\alpha\to-\infty$ can be used, in combination with the opposite  $\alpha\to+\infty$ limit, to estimate the second-order correlation energy $E\c\glt[\rho ]$ without using virtual orbitals. The same procedure works very well also for the Be atom, but breaks down in the case of Ne$^{6+}$, because of strong near-degeneracy effects.

The analysis carried out here extends our knowledge on the exact properties of the exchange-correlation functional $E\xc[\rho]$, and can be used to test and improve approximations. Another possible application of the present results could arise in the framework of a recently proposed approach to the many-electron problem that contains a mixture of Hartree-Fock and Hartee-Fock-Bogoliubov methods \cite{TsuScu-JCP-09}. This approach is based on the splitting of the Coulomb electron-electron interaction $1/r$ as $-1/r+2/r$. The attractive part $-1/r$ triggers the  Hartee-Fock-Bogoliubov solution. An energy density functional to describe dynamical correlation is also added to the method \cite{TsuScu-JCP-09}. For this latter point, the extension to negative coupling $\alpha$ of the adiabatic connection could prove useful in defining and constructing the appropriate energy functional needed in the method.

\section*{Acknowledgments}
We thank M. Pindl for the CI energy of the attractive electron cluster with $N=4$, and A. Savin for useful discussions and for the data of Ref.~\onlinecite{ColSav-JCP-99}.

\appendix

\section{Two electrons on the surface of a sphere}
\label{appElSp}

`Sp2' is a system of two electrons that are confined to the 2D surface ${\cal S}_2$
of a sphere with given radius $R$. Recently \cite{Sei-PRA-07,LooGil-PRA-09,LooGil-PRL-09}, this system has been
studied extensively. It is of particular interest for DFT, since its 2D ground-state density
for all values of the interaction strength $\alpha\in{\sf R}$ is distributed uniformly
over the spherical surface. Consequently, the external potential $v\ext^\alpha([\rho];\rv)$
in the Hamiltonian \eqref{Hv} is trivial and known from the beginning.
The 2D confinement of these electrons, however, gives rise to an unusually large value \cite{Sei-PRA-07,LooGil-PRA-09} of
\beq
E\c\glt(R)=(4\ln2-3)\;{\rm Ha}=-227.4\;{\rm mHa}.
\label{log}
\eeq
This effect can be traced (see Fig.~\ref{fig_Sp2Sp3}) to the strongly negative energy of a 2D AEC \cite{Sei-PRA-07},
\beq
E_{-1}^{N=2}=
\left\{\begin{array}{r}-1\;{\rm Ha}\quad{\rm (2D),} \\ -\frac14\;{\rm Ha}\quad{\rm (3D).} \end{array}\right.
\eeq
While the value \eqref{log} is independent of the spherical radius $R$, the remaining functionals of Table~\ref{tab_EcGL2} are
\begin{eqnarray}
E\x(R)&=&-\frac{e^2}R\;\equiv\;-\frac12U(R),\nonumber\\
W_\infty(R)&=&-\frac32\frac{e^2}R,\nonumber\\
W'_\infty(R)&=&\frac14\Big(\frac{a_B}{R}\Big)^{3/2}\frac{e^2}{a_B},
\end{eqnarray}
with $R=a_B$ in Table~\ref{tab_EcGL2}.

A similar, but more realistic two-electron system is `Sp3' which has its electrons
confined to the 3D surface ${\cal S}_3$ of a sphere in 4D space,
\beq
{\cal S}_3=\Big\{(x,y,z,u)\in{\sf R}^4\,\Big|\,x^2+y^2+z^2+u^2=R^2\Big\}.
\eeq
Again, the ground-state density is distributed uniformly, but now in a 3D finite
(curved) space ${\cal S}_3$. Consequently, we obtain a value very close the one for the He atom (see Table~\ref{tab_EcGL2}) \cite{Sei-UNPUB-XX,LooGil-PRL-09}
   \beq
E\c\glt(R)=-47.6\;{\rm mHa}.
\eeq
The remaining functionals of Table~\ref{tab_EcGL2} are \cite{Sei-UNPUB-XX},
\begin{eqnarray}
E\x(R)&=&-\frac8{3\pi}\frac{e^2}R\;\equiv\;-\frac12U(R),\nonumber\\
W_\infty(R)&=&-\left(\frac{1}{2}-\frac{16}{3\pi}\right)\frac{e^2}R,\nonumber\\
W'_\infty(R)&=&\frac{3}{8}\Big(\frac{a_B}{R}\Big)^{3/2}\frac{e^2}{a_B},
\end{eqnarray}
with $R=a_B$ in Table~\ref{tab_EcGL2}.

\end{document}